\documentclass[a4paper,11pt]{article}
\pdfoutput=1 

\usepackage{jcappub} 

\usepackage[T1]{fontenc} 
\usepackage{latexsym}
\usepackage{amssymb}
\usepackage{epsfig}
\usepackage{amsmath}
\usepackage{float}
\usepackage{color}
\usepackage{enumitem}
\usepackage{placeins}
\usepackage{hhline}
\usepackage{graphicx}
\usepackage{subcaption}
\usepackage{lscape}

\newcommand{\be}{\begin{equation}}
\newcommand{\ee}{\end{equation}}
\newcommand{\bea}{\begin{eqnarray}}
\newcommand{\eea}{\end{eqnarray}}

\newcommand{\Fb}{\bar{F}}
\newcommand{\Zb}{\bar{Z}}
\newcommand{\rhob}{\bar{\rho}}
\newcommand{\Xb}{\bar{X}}
\newcommand{\Pb}{\bar{P}}
\newcommand{\phib}{\bar{\phi}}
\newcommand{\adotoa}{\mathcal{H}}

\newcommand{\MontePython}{\textsc{MontePython}}
\newcommand{\CLASS}{\textsc{class} }

\begin{document}

\title{Momentum transfer models of interacting dark energy}

\author{Mark S. Linton,$^{a}$}
\author{Robert Crittenden,$^{a}$}
\author{Alkistis Pourtsidou$^{b,c,d,e}$}
\affiliation{
$^a$ Institute of Cosmology and Gravitation, University of Portsmouth,\\
 Dennis Sciama Building, Burnaby Road, Portsmouth, PO1 3FX, United Kingdom \\
$^b$ Higgs Centre for Theoretical Physics, School of Physics and Astronomy,
\\  The University of Edinburgh, Edinburgh EH9 3FD, UK \\
$^c$ Institute for Astronomy, The University of Edinburgh, Royal Observatory, \\ Edinburgh EH9 3HJ, UK \\
$^d$ School of Physics and Astronomy, Queen Mary University of London, \\Mile End Road, London E1 4NS, United Kingdom \\
$^e$ Department of Physics and Astronomy, University of the Western Cape,\\ Cape Town 7535, South Africa
}
\emailAdd{mark.linton@port.ac.uk}
\emailAdd{robert.crittenden@port.ac.uk}
\emailAdd{alkistis.pourtsidou@ed.ac.uk}

\begin{abstract}{
We consider two models of interacting dark energy, both of which interact only through momentum exchange. One is a phenomenological one-parameter extension to $w$CDM, and the other is a coupled quintessence model described by a Lagrangian formalism. Using a variety of high and low redshift data sets, we perform a global fitting of cosmological parameters and compare to $\Lambda$CDM, uncoupled quintessence, and $w$CDM. We find that the models are competitive with $\Lambda$CDM, even obtaining a better fit when certain data sets are included.
}
\end{abstract}

\maketitle

\section{\label{sec:Intro}Introduction}
The nature of dark energy (DE) remains an open and intriguing question of modern cosmology. Whether the observed accelerated expansion of the Universe is due to a cosmological constant, a dynamical field like quintessence (see \cite{Copeland:2006wr} for a review), a signature of modified gravity (see \cite{CliftonEtal2011} for a review), or some more exotic or undiscovered phenomena, is still a matter of debate. 

As current and forthcoming high precision experiments explore the universe using multiple probes and high precision measurements, we are able to test our cosmological models with greater rigour than ever before. At the moment, the standard cosmological model, $\Lambda$CDM, is still the statistically preferred model to fit CMB and LSS data \cite{Aghanim:2018eyx, Anderson_2012, Song:2015oza, Beutler_2016, Hildebrandt:2016iqg, Abbott:2017wau, Troster:2020kai,Heymans:2020gsg}, but the existence of the Hubble constant ($H_0$) and structure growth ($\sigma_8$) cosmological tensions has fueled interest in  a plethora of alternative models to see if they can alleviate them (see e.g. \cite{Joudaki:2016kym, SpurioMancini:2019rxy, DiValentino:2019jae,Troster:2020kai,Knox:2019rjx,DiValentino:2020vvd}). 

Models of interacting dark energy (IDE) are more popular than ever. This broad range of models does away with the assumption in $\Lambda$CDM that the DE and dark matter (DM) are isolated systems and allows them to interact (see, e.g., \cite{Amendola:1999er,Pourtsidou:2013nha, Tamanini:2015iia, DiValentino:2019jae}). Additionally these models appear to be flexible enough to solve perceived problems in cosmology such as the $H_0$ or the $\sigma_8$ tensions \cite{Pourtsidou:2016ico,DiValentino:2017iww, Martinelli:2019dau}, although it is not clear if they can truly restore concordance \cite{Knox:2019rjx,Gomez-Valent:2020mqn,Efstathiou:2021ocp}. Many of these models come with undesirable features such as complex or large quantum corrections \cite{DAmico:2016jbm, Marsh:2016ynw}, unnaturally small coupling parameters and new, ad hoc degrees of freedom. The most common interacting dark energy models involve energy exchange in the background, which has severe observational consequences and fails to fit CMB data \cite{Bean_2008,Xia:2009zzb,Amendola_2012,Gomez-Valent:2020mqn}. 
Models that involve momentum-only interactions can circumvent many of these problems. These models leave the background evolution of the Universe unchanged, 
only affecting the perturbations \cite{Simpson:2010vh, Pourtsidou:2013nha, Skordis:2015yra, Linton:2017ged, Chamings:2019kcl,Kase:2019veo,Kase:2019mox}. This allows them to fit the CMB data very well, and to alleviate the $\sigma_8$ tension \cite{Pourtsidou:2016ico,Baldi:2016zom,Jimenez:2021ybe}. 
In this work we test two such models using data from the CMB, SNe Ia, BAO, and cluster counts. 

The paper is structured in the following way: in \autoref{sec:models} we describe the two pure momentum transfer models under consideration and outline their key features. In \autoref{sec:MCMC} we discuss the methodology of the parameter fitting, the data and surveys used, and outline our results. We discuss the impact on the tensions in \autoref{sec:disc} and conclude in  \autoref{sec:concl}.

\section{\label{sec:models}The models}
As we already mentioned, both of the models we will examine in this paper have interactions that exhibit pure momentum exchange at the level of perturbations, and no energy exchange in the background. The first is an elastic scattering (ES) model, first proposed in \cite{Simpson:2010vh}. The second is part of the so-called \emph{Type 3} (T3) class of models constructed in \cite{Pourtsidou:2013nha}, and a particular subclass where the sound speed $c_s^2 \neq 1$ as discussed in \cite{Linton:2017ged}. For both models we have modified a version of \CLASS \cite{Lesgourgues:2011re,Blas:2011rf} and will be using the Monte Carlo code \MontePython  \cite{Audren:2012wb, Brinckmann:2018cvx} for cosmological parameter estimation.
Both models have shown the ability to suppress late time structure growth when compared to a $\Lambda$CDM universe \cite{Pourtsidou:2016ico, Bose:2016qun, Baldi:2016zom,Linton:2017ged,Chamings:2019kcl}.

\subsection{\label{sec:ES}Elastic Scattering}

This phenomenological elastic scattering model adds a momentum coupling to the $w$CDM model inspired by Thompson scattering \cite{Simpson:2010vh}. $w$CDM is an extension of $\Lambda$CDM but DE is treated as a fluid where the equation of state $w\neq-1$. It is one of the most popular extended models and a priority model to be tested with Stage IV cosmological surveys like Euclid \cite{Amendola_2012,Blanchard:2019oqi}. The ES model introduces a single new parameter when compared to $w$CDM, $\sigma_D$, that is defined as the scattering cross section between DE and DM. 

To understand how this new parameter affects the cosmology it is useful to initially look at the perturbation equations for the standard (uncoupled) $w$CDM model.
Working in the Newtonian gauge, the dark matter perturbation equations in $w$CDM are:
\bea
\delta'_{\text{DM}}=& -\theta_{\text{DM}}+\frac{h'}{2},\\
\theta'_{\text{DM}}=& -\mathcal{H} \theta_{\text{DM}} \, ,
\eea
where $\delta$ is the density contrast and  $\theta$ is the velocity divergence, $\mathcal{H}$ the conformal Hubble parameter, $h$ is the metric perturbation and prime refers to a derivative with respect to conformal time \cite{Ballesteros:2010ks}.

We can now see how the introduction of the scattering cross section modifies the second of these perturbation equations  \cite{Simpson:2010vh,Baldi:2016zom}:  
\be \label{eq:ES_def}
\theta'_{\text{DM}}= A\Delta \theta-\mathcal{H}\theta_{\text{DM}} \, ,
\ee
where $\Delta \theta:= \theta_{\text{DM}}-\theta_{\text{DE}}$, and,
\be
A:=(1+w)\sigma_D \frac{\rho_{\text{DE}}}{\rho_{\text{DM}}}n_{\text{DM}} \, .
\ee
Here, $n_{\text{DM}}$ is the number density of dark matter (DM).
As the value of $n_{\text{DM}}$ remains unconstrained, we choose to rewrite $A$ in the following way:
\be
A=(1+w)\xi \rho_{\text{DE}} \, ,
\ee
where $\xi=\sigma_D\frac{n_{\text{DM}}}{\rho_{\text{DM}}}=\frac{\sigma_D}{m_{\text{DM}}}$. In principle, $\xi$ has to be positive as it is a ratio of a cross-section and a mass; however, we will also consider couplings with a negative sign. When we use a negative coupling it can be thought of as a redefinition of the model, taking equation \autoref{eq:ES_def} and placing a minus sign in front of $A$. For simplicity we will just notate this as a coupling with a negative sign.

\begin{figure}[H]
  \centering
  \includegraphics[width=1\textwidth]{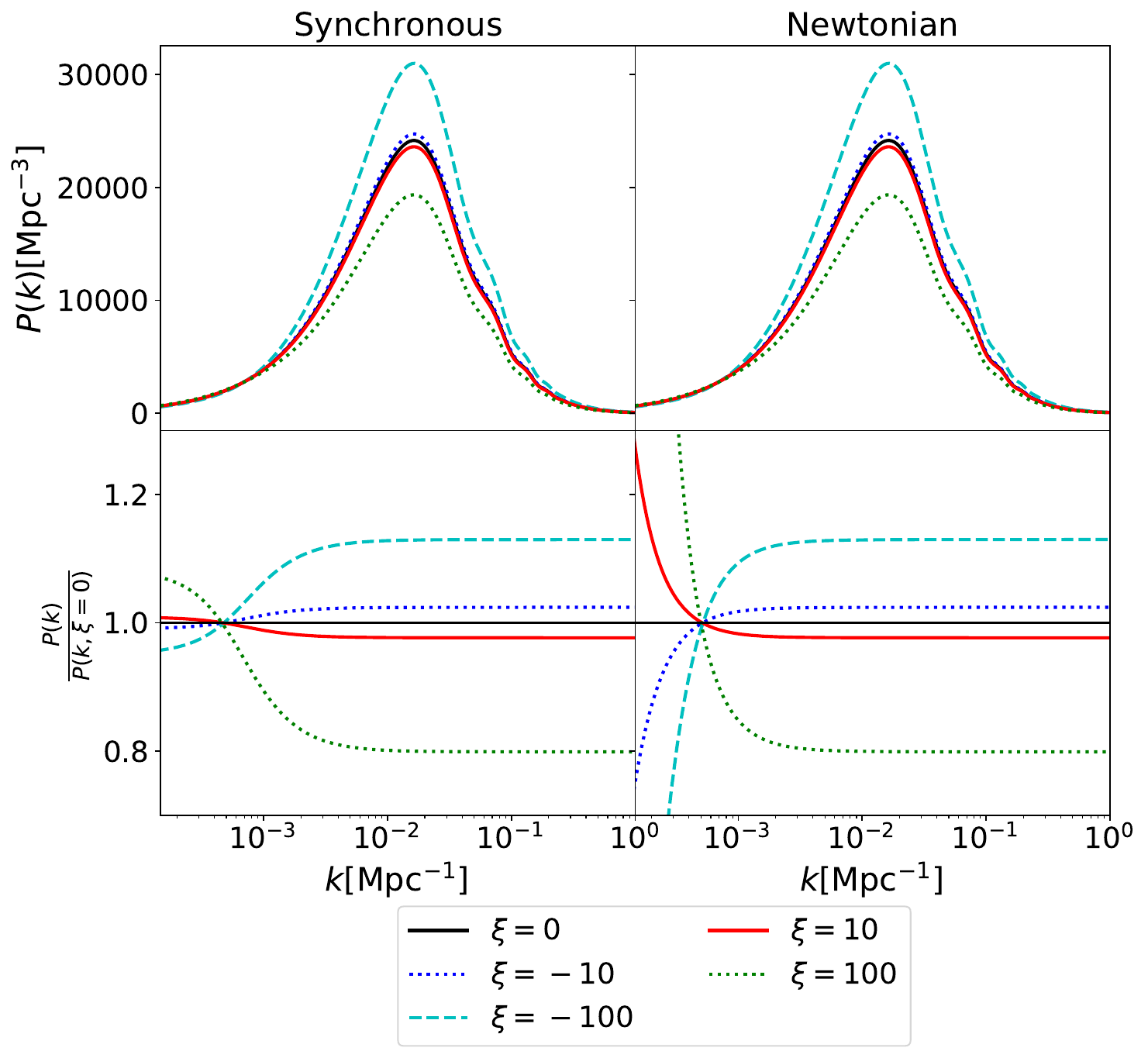}
  \caption{The plots show the effects of the elastic scattering model described in \cite{Simpson:2010vh}. The matter power spectrum (top) and its relative suppression with respect to $w$CDM (bottom) are shown for different values of $\xi$, and $w=-0.9$. The differences between the synchronous (left) and Newtonian (right) gauges are also shown.} 
\label{fig:Matter_power}
\end{figure}

Throughout  this paper we will work in the synchronous gauge unless otherwise stated. Some previous work on the ES model has been done in the Newtonian gauge \cite{Baldi:2016zom,Bose:2017jjx}, however any true observable is gauge independent \cite{Bruni:2011ta}. Although the matter power spectrum is not a true observable, we can see that the two gauges agree on all but the largest scales as demonstrated in \autoref{fig:Matter_power}. Additionally the coupling term $A \Delta \theta$ is gauge independent.

\begin{figure}[H]
  \centering
  \includegraphics[width=0.7\textwidth]{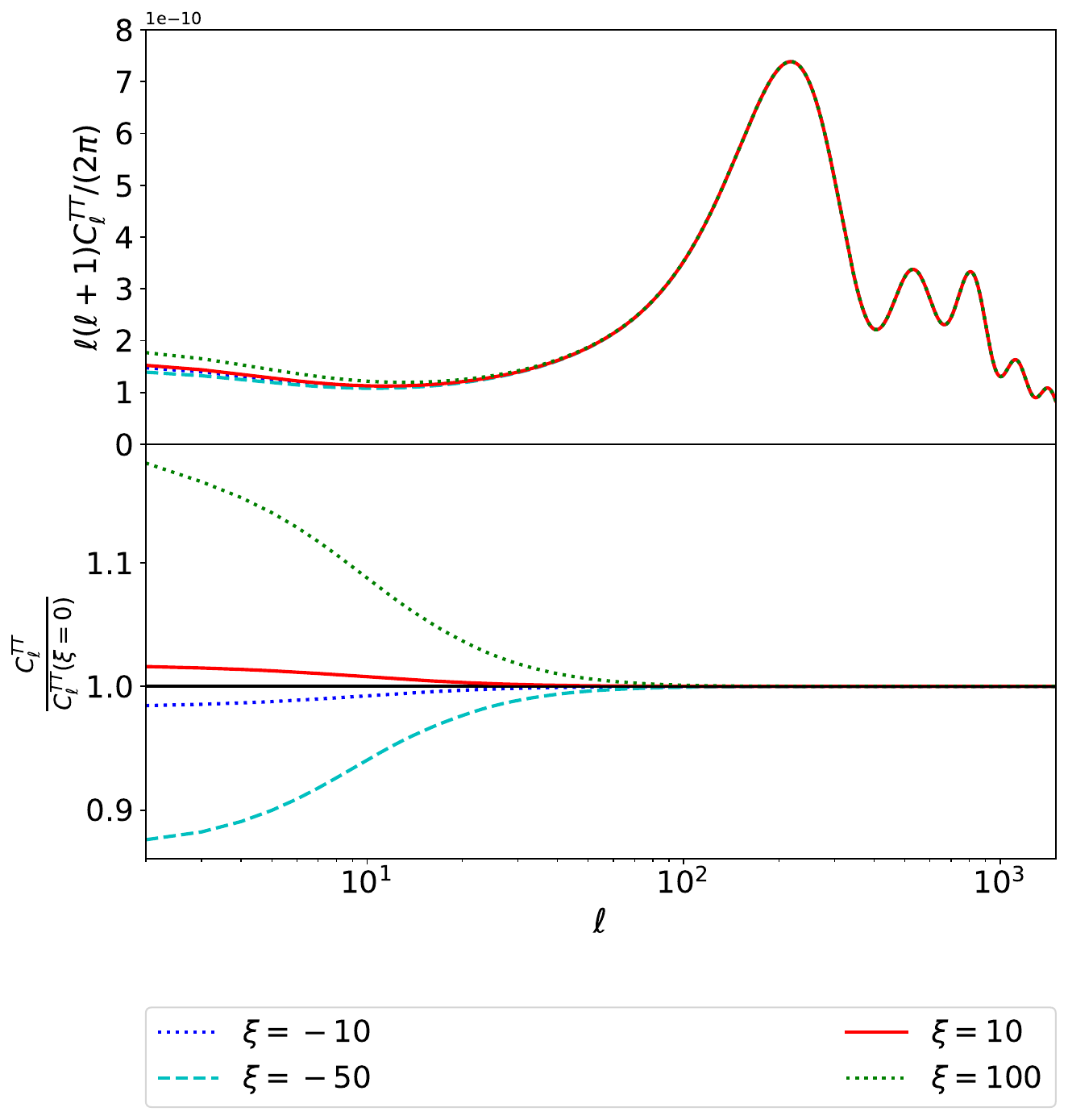}
  \caption{ The CMB temperature-temperature (TT) power spectra for a range of values of $\xi$.
The plots show the predictions from the ES models and uncoupled $w$CDM model (upper), 
the ratio between the coupled models and uncoupled (lower) with $w=-0.9$. } 
\label{fig:CMB_xi}
\end{figure}

Like other momentum-only models of IDE, the effect these models have on the CMB is minimal when compared to an equivalent uncoupled model. In \autoref{fig:CMB_xi}, we see that these models only affect the CMB on the largest scales. This is due to a modification to how the perturbations and potential wells evolve, leading to the well-known late-time integrated Sachs-Wolfe (ISW) effect \cite{Sachs:1967er}.

One of the key differences of the Elastic Scattering model compared to the \emph{Type 3} model (discussed below) is the fact that in the $\xi \rightarrow 0$ (or $w\rightarrow -1$) limit we recover $w$CDM. This relates to another feature of the model that must be taken into account: there is a non-trivial relation between $A$, $\xi$ and $w$. While $\xi$ can be very large, if $w+1$ is very close to 0, then its effect will be very small. This becomes an even greater issue if one considers cosmologies where $w<-1$ is allowed, as this will cause the sign of $A$ to flip.
Despite this, it is still interesting to explore the behaviour of the model when $w<-1$; however we will examine the $w<-1$ and $w>-1$ branches separately to help  disentangle the $\xi$ and $w$ dependence. 

\begin{figure}[H]
  \centering
  \includegraphics[width=0.7\textwidth]{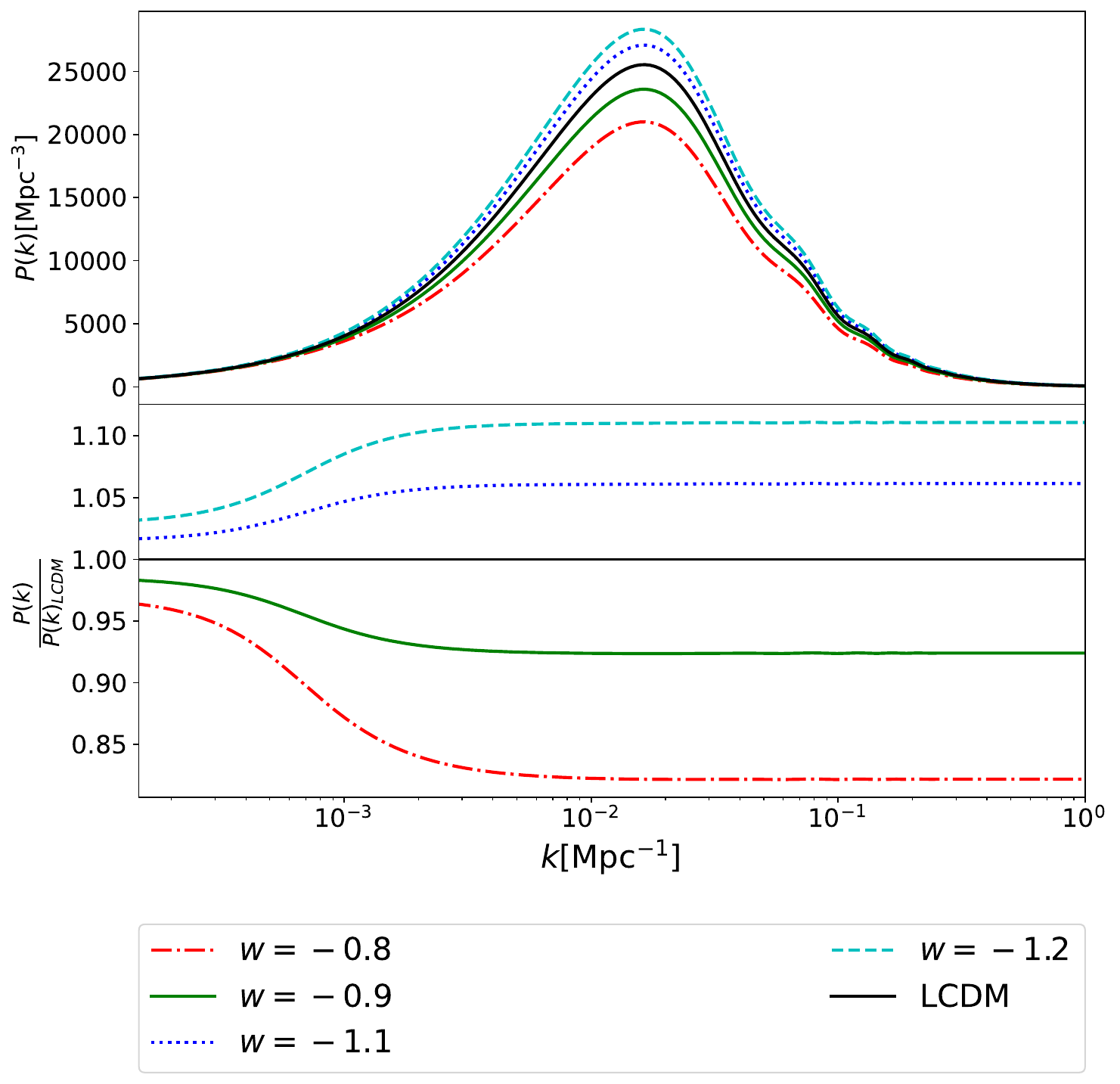}
  \caption{A comparison of the synchronous gauge, linear matter power spectrum $P(k)$ at $z = 0$ for a range of values of  $w$ and fixed coupling, $\xi=10$ (top).   The lower plot shows the ratios between these ES models and $\Lambda$CDM.
 } 
\label{fig:w_0_pk}
\end{figure}

\begin{figure}[H]
  \centering
  \includegraphics[width=1.0\textwidth]{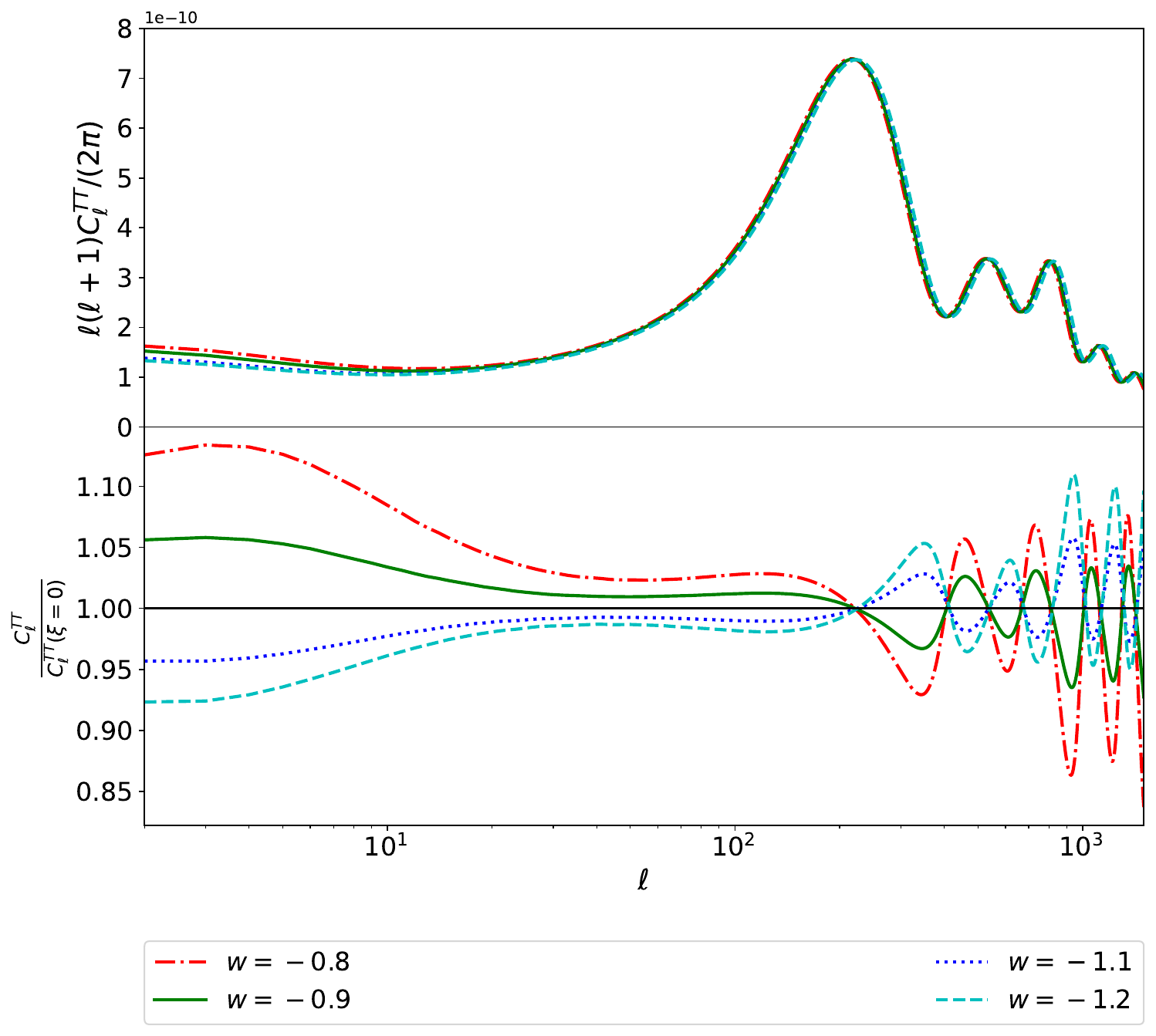}
  \caption{Comparing the CMB temperature (TT) power spectra for a range of values of $w$.
The plots show the predictions from the ES model, with fixed coupling $\xi=10$ for different values of $w$ (top) as well as
the ratio between these models and the $\Lambda$CDM case (bottom).} 
\label{fig:w0_CMB}
\end{figure}

In addition, it is interesting to examine the observational effects of $w$ and $\xi$ in more detail to understand if and how the effects can be separated. A comparison of the effects of $w$ on the matter power spectrum can be seen in \autoref{fig:w_0_pk}. Comparing this and \autoref{fig:Matter_power}, which shows the effect on the matter power spectrum for different values of $\xi$, one might be concerned that the two effects are so entangled that it may not be possible to separate them.  However, \autoref{fig:CMB_xi} shows how different values of $\xi$ affect the CMB temperature power spectrum; one can see that the effect is only visible as a change to the integrated Sachs-Wolfe effect. Whereas in \autoref{fig:w0_CMB} the value of $w$ is varied, here one can see the change in the peaks is of the order of $5 \%$.  The key difference is that $w$, unlike the coupling $\xi$, affects the background evolution and thus the distance to the last-scattering surface, leading to a shift in the angular positions of the Doppler peaks.

\subsection{\label{sec:type3} \emph {Type 3} models }

While the elastic scattering model is a phenomenological implementation of  pure momentum transfer interacting dark energy, similar models can be constructed from a Lagrangian formalism \cite{Pourtsidou:2013nha, Tamanini:2015iia}. 
In this approach, it can be shown that momentum-only exchange arises in so-called \emph{Type 3} models where the Lagrangian takes the form,
\begin{equation}
L = F(Y,Z,\phi)+f(n).  
\end{equation} 
Here $\phi$ is the dark energy scalar field, $n$ is the dark matter fluid number density, $Y \equiv \frac{1}{2}(\nabla_\mu \phi)^2$ is the usual kinetic term and  $Z \equiv u^\mu \nabla_\mu \phi$.  It is this latter term that couples the dark matter fluid velocity, $u^\mu$,  to the gradient of the scalar field. 
Here, the coupling current is,
\be
J_{\mu} = q^\beta_{\; \mu} \Big(\nabla_\nu (F_Z u^\nu)\nabla_\beta \phi 
+ F_Z \nabla_\beta Z + Z F_Z u^\nu \nabla_\nu u_\beta \Big) \, ,
\ee 
with, 
$q^\nu_{\; \mu} =  u^\nu u_\mu +\delta^\nu_{\; \mu}$ and the subscripts denote derivatives, e.g.  $F_Z=d F/d Z$. 
From the above formula we find $J_0 = 0$ up to second order, but $\delta J_i \neq 0$, hence \emph{Type 3} is a theory of pure momentum exchange up to linear order.

Different implementations of these models have previously been studied \cite{Pourtsidou:2013nha, Skordis:2015yra, Pourtsidou:2016ico, Linton:2017ged}.   Here we focus on a specific model with variable sound speed ($c_s^2$), which was first presented in \cite{Linton:2017ged}. The Lagrangian takes the following form:
\begin{equation}
L = Y+\beta_1 Z^3+V(\phi) \, . 
\end{equation} 
Working in the DM frame, the Lagrangian is 
\be 
L=\frac{1}{2}\left(1+2 \beta_1 \frac{\dot{\phi}}{a}\right)\dot{\phi}^2-\frac{1}{2}|\vec{\nabla} \phi|^2 -V(\phi) \, ,
\ee
where $\beta_1$ is a coupling constant with dimensions $\bigg[\frac{1}{\dot{\phi}} \bigg]$.
The dark energy originates from a quintessence field assumed to have an exponential potential 
\begin{equation}
V(\phi)=V_0 e^{-\lambda \phi} \, ,
\end{equation}
where $\lambda$ and $V_0$ are constants. 

\subsection{Comparing the models}

As mentioned above, these models are both IDE models and, to linear order, only effect the perturbations through momentum exchange.
In \cite{Skordis:2015yra} the authors compared the ES model from \cite{Simpson:2010vh}, with a \emph{Type 3} model. The authors followed and extended the parameterized post-Friedmannian approach, previously applied to modified gravity theories \cite{Baker:2011jy}, in order to include interacting dark energy. This was then used to describe a very broad range of scalar field DE models that are coupled to a DM fluid.
For a Type 3 model the momentum flux, $S$, is \cite{Skordis:2015yra}
\be 
S= B_3 \delta_{DE}+B_5 \theta_{DE}+B_6 \theta_{CDM}, 
\ee
where the $B_3$, $B_5$ and $B_6$ coefficients depend on quantities  that appear in the Lagrangian and their derivatives.

In the case of the ES model,
\be
S=(1+w)\rho_{\rm DE}\xi \Delta \theta,
\ee
so for the models to be equivalent, we would require 
\begin{align}
B_3=&0, \\
B_5=&-B_6=(1+w)\rho_{\rm DE}\xi.
\end{align}
If we examine the expressions for $B_3$, $B_5$ and $B_6$ for a Type 3 model, we get
\begin{align}
B_3  &= \frac{1}{1-\frac{\Zb \Fb_Z}{\rhob_{DM}}}  \frac{\Zb \Fb_Z c_s^2}{1+w} \, ,
\\
B_5 &=
 \frac{a}{1-\frac{\Zb \Fb_Z}{\rhob_{DM}}}
  \bigg[
 \Xb \left( \frac{  \Fb_Z }{\Fb_Y} - \Zb \right)
+  \Fb_Z   \left[ \frac{\mu}{a\Fb_Z} - \frac{F_\phi}{F_Y} \right] 
\bigg] \, ,
\\
B_6 &= - B_5 +3\mathcal{H}(1+w) B_3 \, ,
\label{type_3_coeff}
\end{align}
where, 
\begin{align*}
 \mu \equiv & \frac{3 F_Z}{\Zb \Fb_Y} (c_s^2 - c_a^2) (\rhob_{DE} + \Pb_{DE})\mathcal{H} \, ,  \\
\Xb  \equiv & \frac{1}{a} \left[ (\Zb \Fb_{ZY} - \Fb_{ZZ})\dot{\Zb} -  \Fb_{Z\phi} \dot{\phib} -3\mathcal{H}   \Fb_Z \right] \, , \\
 \dot{\Zb}  =&- 3\adotoa  \Zb \left[ c_a^2  + \frac{aF_\phi}{3\adotoa(  \Zb \Fb_Y - \Fb_Z  )} \right] \, .
\end{align*}
The adiabatic sound speed, $c_a^2$, is given by 
\begin{eqnarray}
 c_a^2 &=& 
 \frac{ 3  \adotoa  ( \Zb F_Y  - F_Z)  - a \left[ F_\phi +  \Zb^2 F_{Y\phi} - \Zb F_{Z\phi} \right]
}{3\adotoa \Zb (\Fb_Y  + 2 \Zb \Fb_{YZ} - \Zb^2 \Fb_{YY} - F_{ZZ})} -\frac{aF_\phi}{3\adotoa(  \Zb \Fb_Y - \Fb_Z  )} \, ,
\end{eqnarray}
and the sound speed, $c_s^2$, is
\begin{equation}
c^2_s=\frac{\Zb\Fb_Y-\Fb_Z}{\Zb\left(\Fb_Y+2\Zb\Fb_{YZ}-\Fb_{ZZ}-\Zb^2\Fb_{YY}\right)} \, .
\end{equation}
Although this is a complex set of equations, we can see that in the limit that $c_s^2 \rightarrow 0$, $B_3 \rightarrow 0$ and $B_5=-B_6$ reducing the \emph{Type 3} model to something that appears similar to the ES model.

However, finding a Lagrangian where $c_s^2 \rightarrow 0$ is challenging. It is shown in \cite{Linton:2017ged} that for an action of the form 
\be \label{eq:simple_T3}
F=Y+\beta_{n-2}Z^n+V(\phi) \, ,
\ee
when $n>2$, the minimum value of $c_s^2=\frac{1}{n-1}$. So $c_s^2 \rightarrow 0$, only in the case where $n \rightarrow \infty$. However, it is unclear how physical a model with a very large $n$ is.  In addition, it has also been shown \cite{Chamings:2019kcl} that for a model in this form $\dot{\bar{\phi}}$ is inversely proportional to $n$, so as $n$ increases, the field becomes less dynamic.

Although finding a model where $B_3=0$ and $B_5=-B_6$ appears difficult, this does not mean that the models are dissimilar. It is important to consider the effect these parameters have on observables. In the coupling, $B_3$ is multiplied by $\delta_{DE}$ which is shown in \cite{Linton:2017ged} to be very small for the Type 3 model considered here. Based on this, it seems reasonable to assume that this term will be small compared to the other coupling terms and that the contribution of $B_3 \delta_{DE}$ to $S$ and any observables will be minimal, compared to contributions from the other terms.

It is difficult to compare the size of the two contributions to $B_6$ in Eq. \ref{type_3_coeff}. The complexity of the two functions means it is challenging to do an order of magnitude calculation to estimate their relative size. Using a Boltzmann solver such as \CLASS one can test the size of these terms, but we have found the results inconclusive suggesting that the magnitude $B_5$ and $B_6$ may be similar at some scales but different at others.

Although the construction of the two models under consideration in this work is different, their main common  feature is that the interaction is  pure momentum exchange that only affects perturbations. We find that this leads to similar results in our MCMC analysis, but as we will see, some important differences remain.

It is also useful to note that there exist IDE models with interesting observational consequences that do include energy transfer in the background. For example, \cite{Shafieloo:2016bpk} considers metastable dark energy with radioactive-like decay, where the coupling is a constant depending on the intrinsic dark energy properties. In terms of the observables, the model of \cite{Shafieloo:2016bpk} has significant effects on the CMB, as well as on the expansion history of the Universe. We also note that a Bayesian analysis in \cite{Shafieloo:2016bpk} finds that a sub-class of these models in which dark energy decays into dark matter leads to lower values of the Hubble parameter at large redshifts relative to $\Lambda$CDM.

\section{MCMC analysis}
\label{sec:MCMC} 

To constrain our models of IDE we perform an MCMC analysis using the  \MontePython{} code \cite{Audren:2012wb,Brinckmann:2018cvx}, and compare with $\Lambda$CDM, $w$CDM and uncoupled quintessence. All of the subsequent MCMC plots were created using {\tt GetDist} \cite{Lewis:2019xzd}. 

\subsection{Data sets}
The data we will consider include\footnote{We have not used more recent BAO and Supernova data (e.g. \cite{BOSS:2016wmc, Pan-STARRS1:2017jku}), but we do not expect our results and conclusions to qualitatively change.}: 
\begin{description}[leftmargin=4em,style=nextline]
\item[CMB:] The $C_\ell^{TT}$-data from Planck 2018~\cite{Aghanim:2018eyx}, including high and low -$\ell$ polarisation as well as the Planck lensing data from Planck 2018~\cite{Aghanim:2018oex}.
\item[BAO:] Baryon Acoustic Oscillation (BAO) data from BOSS~\cite{Anderson:2013zyy}.
\item[JLA:] Joint Light-curve Analysis (JLA) supernova
data~\cite{Betoule:2014frx}.
\item[SZ:] Planck SZ cluster counts~\cite{Aghanim:2018eyx,2014A&A...571A..20P,Ade:2015fva}.
\end{description}

We choose not to include any weak lensing data directly in our MCMC analysis; this is due the lack of knowledge of the non-linear modelling of our chosen IDE models (both the dark matter non-linear modelling and the effects of baryonic feedback). While there has been some work on N-body simulations of the ES model \cite{Baldi:2014ica,Baldi:2016zom}, as well as on perturbation theory predictions \cite{Bose:2017jjx}, to be conservative we will constrain ourselves to the linear regime for both models.
We choose flat priors on the cosmological parameters,
\be
\{ \omega_b,\omega_{cdm},100 \theta_s, \log_{10} A_s, n_s, \tau_{reio} \} \, ,
\ee
and we also include the nuisance parameters required by the Planck and JLA likelihoods. Additionally we consider the derived parameters $H_0$, $\sigma_8$ and $\Omega_m$. 

It is also important to note the relation between $\sigma_8$ and the mass
bias parameter $b$. For the Planck SZ data this is set to $(1-b)=0.62 \pm 0.03$ and is derived from the cluster counts and the CMB \cite{Aghanim:2018eyx}. It also uses
the Tinker et al halo mass function \cite{Tinker:2008ff}, which has been
calibrated against N-body simulations assuming $\Lambda$CDM. This implies a non-linear reanalysis of this mass function is required to ensure consistency, but this is beyond the scope of this paper, and given the non-linear studies of the ES model in \cite{Bose:2017jjx} we do not believe this would have a significant effect. A significantly lower value of this parameter would also solve the $\sigma_8$ tension, however, other observations such as weak lensing, place constraints on this parameter \cite{Aghanim:2018eyx}. We use the SZ likelihood implementation in the \MontePython{} code release, which is based on the Planck 2013 SZ analysis\footnote{T. Brinckmann, private communication.} \cite{2014A&A...571A..20P}.

\subsection{Non-interacting dark energy models}

For context, we briefly describe constraints on non-interacting versions of the models considered here, namely $w$CDM and quintessence.  The CMB data constrains the dark energy behaviour to be close to that of a cosmological constant, but with relatively loose constraints on the dark energy parameters.  Related to this, the $H_0$ peak also broadens; this degeneracy between $w$ and $H_0$ in CMB is well documented \cite{Ade:2015rim}, but it is interesting in the context of the much discussed $H_0$ tension (particularly in the $w<-1$ case) \cite{Bernal:2016gxb, Knox:2019rjx}. 

The dark energy and $H_0$ constraints are tightened considerably by the addition of the BAO and JLA data, making the models very close to $\Lambda$CDM in practice.  
As a result, these non-interacting models also exhibit the $\sigma_8$ tension exhibited by $\Lambda$CDM, which is evident when the Planck SZ is added. Additionally, these uncoupled models have similar $\chi^2$ results to $\Lambda$CDM. In the following, we will see how this tension is moderated in the momentum-transfer models.   

\subsection{Elastic scattering MCMC}

Before beginning our MCMC analysis of the ES model we must first consider the degeneracy highlighted in \autoref{sec:ES}. 
For models where the equation of state is close to $w=-1$, $1+w$ is small and poorly constrained; thus there is an effective degeneracy in the coupling term $A$, between $1+w$ and $\xi$ that prevents convergence of the MCMC.
In order to break this degeneracy, we introduce a new parameter,
\be
\bar{\xi}=\xi(1+w),
\ee
so $A$ can now be written as, 
\be
A=\bar{\xi} \rho_{\text{DE}}.
\ee
We choose a flat prior on $w$ and a logarithmic prior on $\bar{\xi}$. For the $w>-1$ case,
\be
w\in [-1; -0.3],\, \log{\bar{\xi}} \in [-2, 3],
\ee
and for the $w>-1$ case,
\be
w<-1,\, \log{\bar{\xi}} \in [-2, 3].
\ee

\begin{figure}[ht] 
  \centering
  \includegraphics[width=0.9\textwidth]{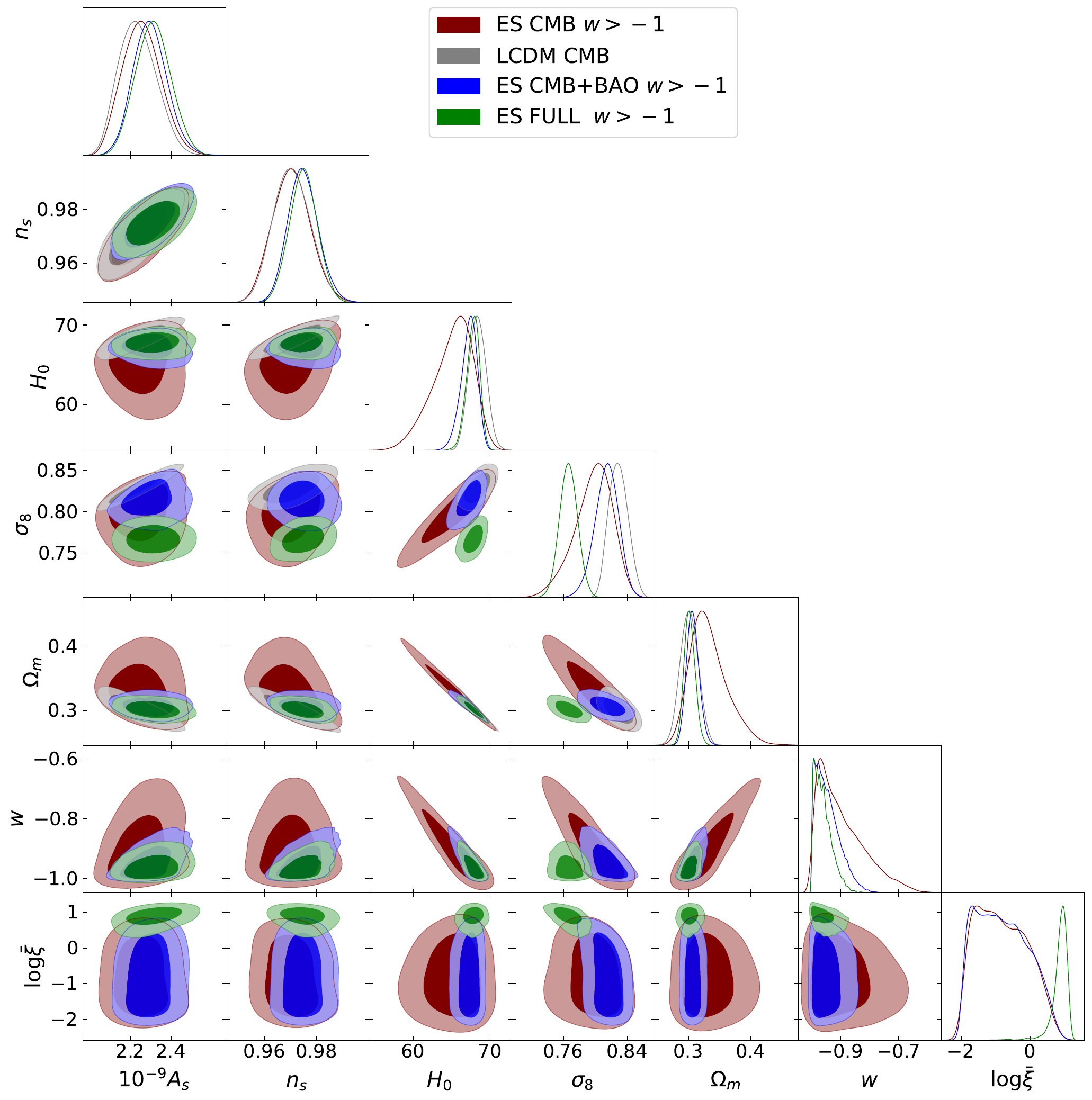}
  \caption{The plot shows the 1 and  $2\sigma$ constraints for $\Lambda$CDM with CMB data (grey) and the ES model $w>-1$  using the CMB data (maroon), using the CMB+BAO (blue) or using CMB, SZ, BAO and JLA (green). With only CMB and CMB+BAO data sets the ES model remains fairly unconstrained, allowing a wide range of values for the parameters $w$ and $\xi$. This leads to large contours on the derived parameters $H_0$ and $\sigma_8$. The inclusion of Planck SZ data independently constrains $\sigma_8$ which can be accommodated in the ES, resulting in a much more tight $\bar{\xi}$ contour.} 
\label{fig:ES_MCMC}
\end{figure}

Examining the ES model with just the CMB data shows some interesting features, as can be seen in \autoref{fig:ES_MCMC}. 
For this data, the model is comparable to the $w$CDM model; 
$w$ and $\bar{\xi}$ are relatively weakly constrained but we see the contours for both $H_0$ and $\sigma_8$ broaden compared to $\Lambda$CDM. 
The weakening of the $\sigma_8$ constraint to admit smaller values is effectively by design; this feature appears in other momentum only IDE models \cite{Pourtsidou:2016ico, Linton:2017ged}  and is consistent with the suppression in the matter power spectrum seen in \autoref{fig:Matter_power} and explored in \cite{Baldi:2016zom, Bose:2017jjx}. It is also a feature of models where $w>-1$, which can be seen in \autoref{fig:w_0_pk}.  Given that the CMB data itself does not constrain $\sigma_8$ directly means that to $2 \sigma$, $\xi$ remains consistent with zero; this was also seen for the models explored in \cite{Pourtsidou:2016ico, Linton:2017ged}.

Adding BAO or supernova data to the CMB, places much stronger constraints on $w$ and prevents this model from accommodating the higher values of $H_0$ observed by distance ladder measurements \cite{Riess:2016jrr}. This is also consistent with previous studies \cite{Knox:2019rjx}. 
Finally, when we add the SZ data we now see that $\xi$ has become much more tightly constrained. We see that the data prefers a small value for $\log{\bar{\xi}}$.  The non-zero coupling allows $\sigma_8$ to be smaller than in the $\Lambda$CDM model, relieving some of the tension seen in that case.  (See below for a further discussion of this tension.) 

\begin{figure}[ht]
  \centering
  \includegraphics[width=0.9\textwidth]{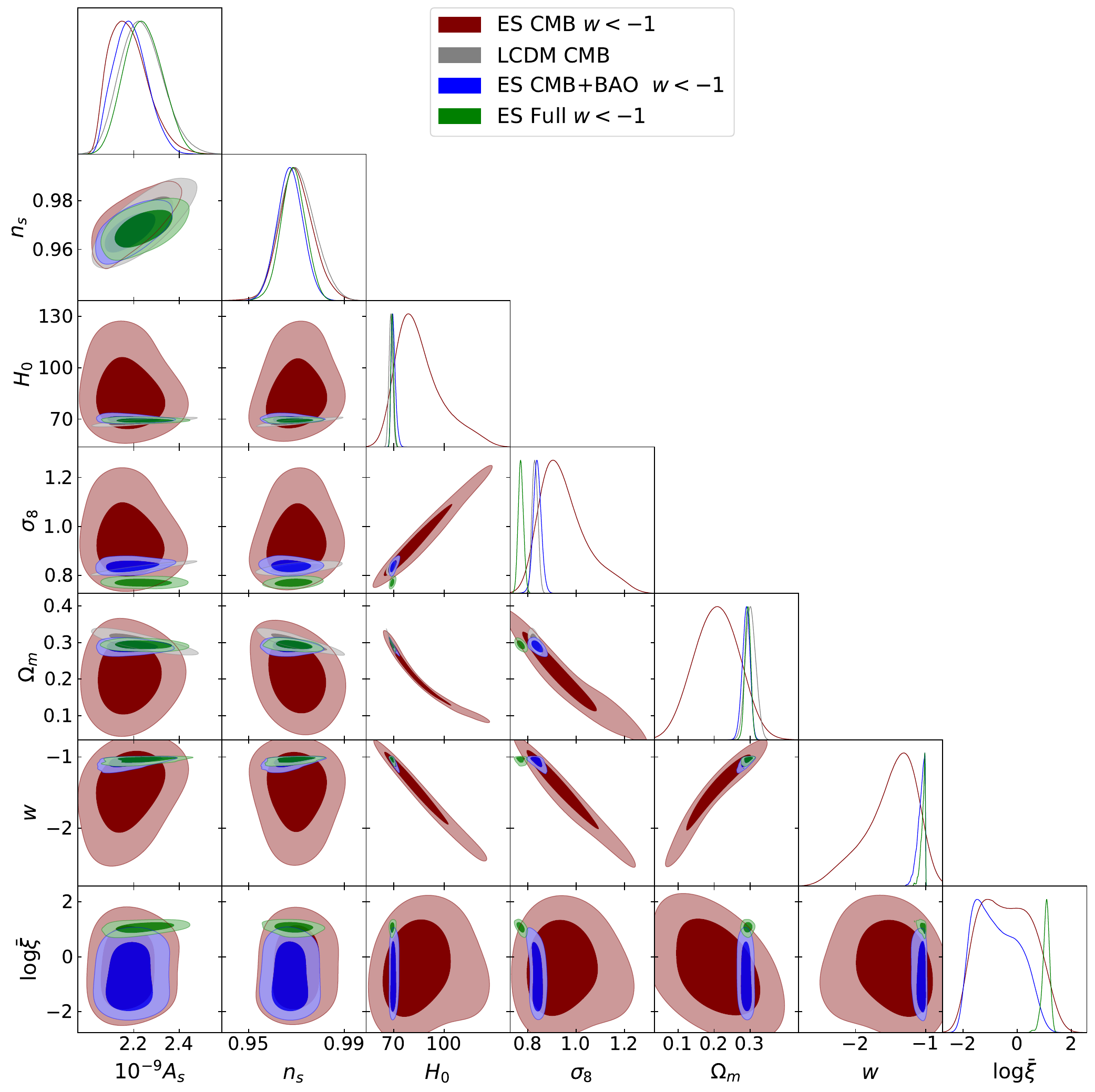}
  \caption{ As above, but with a $w < -1$ ES model.  The plot shows constraints for $\Lambda$CDM with CMB data (grey) and the ES model $w>-1$ using the CMB data (maroon), using the CMB+BAO (blue) and CMB, SZ, BAO and JLA data (green). The results are very similar to the $w>-1$ case seen above because the observations constrain the dark energy equation of state to be close to $w=-1$.}
\label{fig:ES_phant}
\end{figure}

The results for the elastic scattering model with $w<-1$ are shown in  \autoref{fig:ES_phant}. When all the data sets are considered, the $w>-1$ and the $w<-1$ branches look very similar; this is due to the fact that for both cases $w$ is very close to $-1$.  The suppression of $\sigma_8$ indicates a positive value for $A \propto \xi (1+w)$  However, while for $w>-1$ this is equivalent to a positive value for $\xi$, for the $w<-1$ this is actually equivalent to a negative value for $\xi$. 

\subsection{\emph{Type 3} MCMC}

For the \emph{Type 3} models, we exclude the negative values of $\beta_1$ as they lead to pathologies like ghosts \cite{Pourtsidou:2013nha}. Following \cite{Pourtsidou:2016ico, Linton:2017ged} we choose the following priors for $\lambda$ and $\beta_1$:
\be
\lambda  \in [0; 2.1],\, \log{\beta_1} \in [-2, 15] \, .
\ee

For the Type 3 model, $\lambda$ is the parameter that has the biggest influence on the equation of state parameter today, $w$. Similar to what we see in the ES case, the two new parameters, $\lambda$ and $\beta_1$, remain fairly unconstrained when we only consider the CMB data. The lack of constraint on $\lambda$ means that the effective $w$ remains relatively free. It follows that the observational parameters closely related to $w$ and $\beta_1$, $H_0$ and $\sigma_8$, are similarly weakly constrained. In the case of $H_0$ we see similar or lower values when compared to $\Lambda$CDM as we chose to restrict ourselves to the non-phantom branch of these models by having the kinetic term with the usual sign and $\beta_1>0$. With regards to $\sigma_8$, we see that the Type 3 model can accommodate lower values than $\Lambda$CDM.

\begin{figure}[ht]
  \centering
  \includegraphics[width=0.9\textwidth]{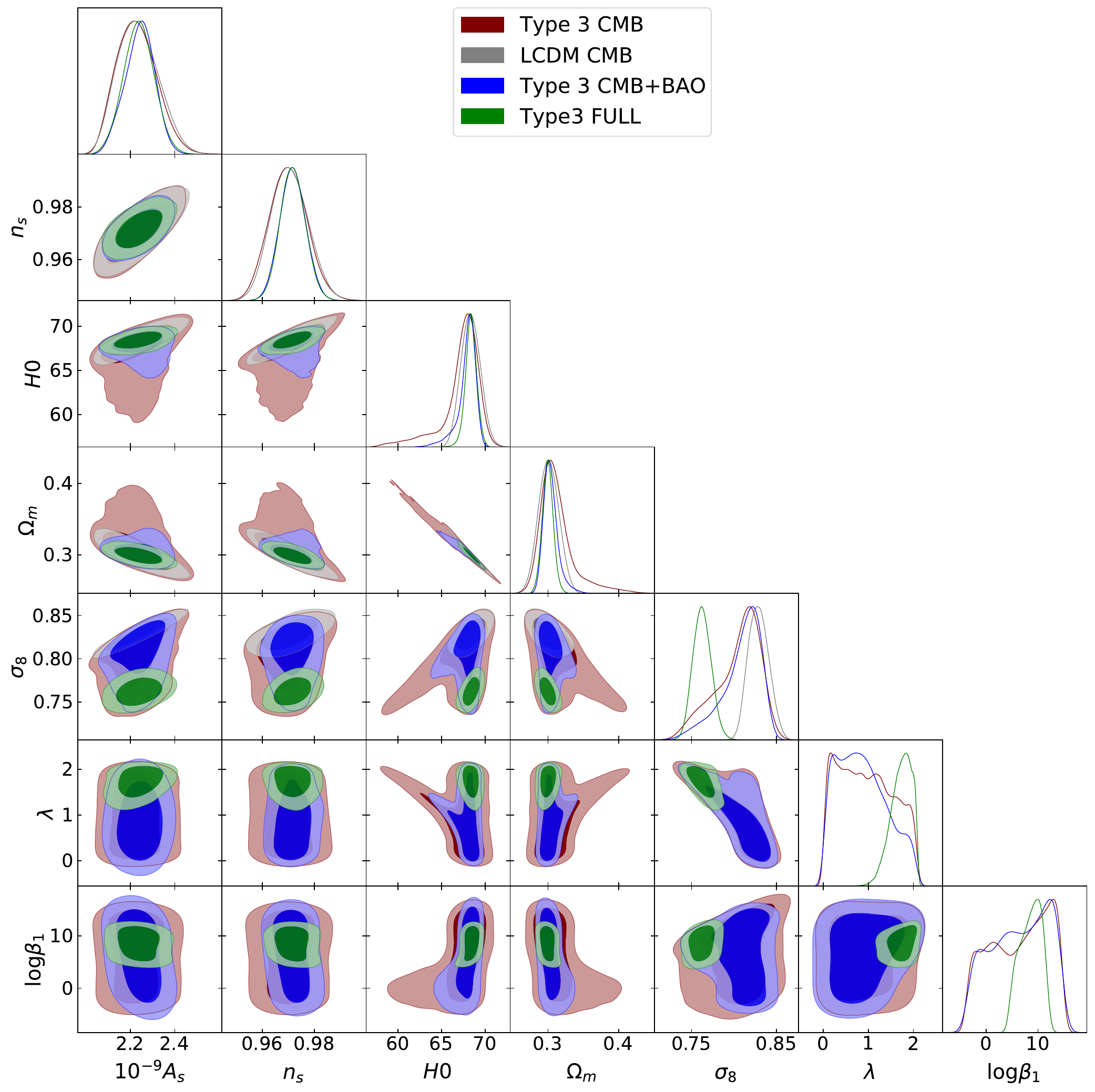}
  \caption{1 and 2 $\sigma$ contours from $\Lambda$CDM  CMB (grey), T3 model using CMB data alone (maroon), using CMB and BAO data (blue), and using CMB, SZ, BAO and JLA data (green).} 
\label{fig:T3_full}
\end{figure}

As we saw in the ES case, introducing the BAO data places tighter constraints on $w$ and $H_0$. For the Type 3 model this means that $\lambda$ is more tightly determined. However, with just the CMB and BAO datasets $\beta_1$ and thus, $\sigma_8$ continue to be unconstrained.

When we add in the Planck SZ data we push $\sigma_8$ down and see that $\beta_1$ settles to a value around $10^8$.  Both the T3 model and the ES model are able to accommodate the lower $\sigma_8$ values arising from the SZ data.   We also see that there is a slight shift of the parameter $\lambda$ when we use all the data compared to the CMB only and the CMB+BAO data.

It is also interesting to compare this model to the Type 3 quadratic case studied in \cite{Pourtsidou:2016ico}. This model has the Lagrangian
\begin{equation}
L = Y+\beta_0 Z^2+V(\phi), 
\end{equation} 
but unlike the previous T3 model, this model does not have a varying sound speed. 
The MCMC analysis shows that these two models are very similar. Both models can fit the data and accommodate lower values of $\sigma_8$ than $\Lambda$CDM. The main difference is the sign and the size of the coupling parameters, $\beta_0$ and $\beta_1$. These changes are a direct result of the different coupling functions, for the model studied in \cite{Pourtsidou:2016ico} $\beta_0 Z^2$ and for the model studied here and in \cite{Linton:2017ged}, $\beta_1 Z^3$. It is useful to consider a dimensionless and dynamical coupling, $\beta_1 Z^3=\beta_0(Z) Z^2$ where $\beta_0(Z)=\beta_1 Z$. For the chosen potential $Z<0$, so for a positive $\beta_1$, $\beta_0(Z)$ is negative, like the quadratic case. This method of looking at the coupling function is also useful when comparing the scale of the coupling. In \cite{Linton:2017ged} the evolution of $\beta_1 Z$ is shown for given parameters, for $\beta_1=10^{15}$, $\beta_0(Z)\approx10^6$ today; this is similar scale to the couplings discussed in \cite{Pourtsidou:2016ico}.

Given that the coupling functions are very similar at late times, the main difference between the models is that one has a varying sound speed and one does not. As the MCMC analysis produces similar results for both models and it suggests that the varying sound speed has a limited effect on observations. It was shown in \cite{Linton:2017ged}, that the cubic model has a sound speed between $1$ and $1/2$. Previous studies have shown that if a change in the sound speed of DE is to produce observational effects, it must have $c^2_s \ll 1$ \cite{Xia:2007km, dePutter:2010vy, Ballesteros:2010ks, Basse:2012wd}. This idea is reinforced by the fact that in \cite{Linton:2017ged}, it is shown that the DE perturbations remain very subdominant when compared to the DM perturbations, despite the reduction in the sound speed.

\subsection{Results}

The best-fit model parameters and the resulting $\chi^2$ values using the CMB and full data sets are shown in  \autoref{table:bf-cmb} and \autoref{table:bf-full} respectively.
With the CMB data alone, we find all of the models fit with comparable $\chi^2$.  The $\Lambda$CDM model is within the $w$CDM and the ES models, and effectively in the Type 3 model if the quintessence potential is made flat enough; the best fit models are all consistent with this and the additional parameters are generally poorly constrained. Thus considering the CMB data alone, $\Lambda$CDM is significantly preferred compared to these more complex models.  

This picture changes considerably when the other data sets are included.  As discussed below, the $\sigma_8$ tension manifests when the Planck SZ data is included, meaning the more complex models provide a significantly improved $\chi^2$.  The contributions to the $\chi^2$ for each data set are shown in \autoref{table:ch2}.

\begin{table}[h]
    \centering
    \begin{tabular}{|l|c|c|c|c|c|} 
\hline 
					&{LCDM } 	&				ES ($w>-1$) &					ES ($w<-1$) & 				Type 3 	\\	
					&CMB		&				CMB &	 					CMB &					CMB		\\ 
\hline
 
Param &  				mean$\pm\sigma$ & 			mean$\pm\sigma$ & 			mean$\pm\sigma$ &			mean$\pm\sigma$ \\
\hline 
$100~\omega_{b }$ & 	$2.234_{-0.027}^{+0.025}$   & 		$2.233_{-0.027}^{+0.025}$&		$2.214^{+0.082}_{-0.18}$&	$2.233_{-0.027}^{+0.026}$	\\
$\omega_{cdm }$   & 	$0.1176_{-0.0023}^{+0.0026}$ & 	$0.1177_{-0.0025}^{+0.0024}$&	$0.124^{+0.068}_{-0.014}$&	$0.1177_{-0.0024}^{+0.0024}$ 	\\
$n_{s }$   & 			$0.9704_{-0.0077}^{+0.007}$ & 	$0.9704_{-0.0077}^{+0.007}$ 	&	$0.966^{+0.020}_{-0.029}$&	$0.9701_{-0.0075}^{+0.0072}$ 	\\
$\tau_{reio }$  & 		$0.0868_{-0.025}^{+0.021}$  & 	$0.0930_{-0.025}^{+0.023}$	&	$0.072^{+0.043}_{-0.035}$&	$0.0875_{-0.025}^{+0.021}$ 	\\
$H_0$   & 				$68.3_{-1.2}^{+1.1}$ & 			$65.1_{-1.8}^{+3.2}$ 	&		$81^{+40}_{-30}$&			$67.25_{-0.85}^{+2.3}$		\\
$10^{9}A_{s }$  & 		$2.226_{-0.1}^{+0.08}$ & 			$2.254_{-0.11}^{+0.089}$ 	&		$2.16^{+0.18}_{-0.15}$&		$2.229_{-0.1}^{+0.081}$		\\ 
$\sigma_8$  & 			$0.828_{-0.013}^{+0.011}$  & 		$0.797_{-0.019}^{+0.028}$ 	&	$0.94^{+0.22}_{-0.14}$&		$0.804_{-0.016}^{+0.034}$	\\
$\Omega_{m }$   & 		$0.302_{-0.015}^{+0.015}$ & 		$0.332^{+0.061}_{-0.051}$	&	$0.27^{+0.58}_{-0.22}$&		$0.312_{-0.027}^{+0.012}$ 	\\
$w_{fld}$&			$-$&							$-0.897_{-0.1}^{+0.028}$	&		$-1.42^{+0.53}_{-0.80}$&		$-$						\\
$w_{scf}$&			$-$&							$-$	&						$-$&						$-0.945_{-0.055}^{+-0.022}$ 	\\
$\log{\bar{\xi}}$&		$-$&							$-0.78$ * 	&		$-0.6^{+1.8}_{-1.5}$&		$-$						\\
$\log{\beta_1}$&		$-$&							$-$	&						$-$&						 $6.31$*		\\
$\lambda$&			$-$&							$-$	&						$-$&						$0.95$* 	\\ 
\hline

$\chi^2$& 				$790.6$&						$791.5$	&					$790$&					$791$					\\
$\Delta \chi^2$&		$-$&							$+0.9$	&					$-0.6$&					$+0.4$						\\
\hline 
 
 \end{tabular} 

 \caption{Constraints of cosmological parameters for $\Lambda$CDM, the T3 model and the ES model obtain using an MCMC and CMB data. When only the CMB data is used $\log{\bar{\xi}}$ for the ES $w>-1$ case, $\log{\beta_1}$ and $\log{\beta_1}$ remain unconstrained and saturate the priors so we do not include standard deviations, these parameters are marked with *.}
 \label{table:bf-cmb}
 \end{table}

\begin{table}[h]
    \centering

\begin{tabular}{|l|c|c|c|c|c|} 
\hline 
					&LCDM &						ES ($w>-1$) & 					ES ($w<-1$) & 					Type 3 				\\	
					&Full&						Full	&						Full&							Full					\\ 
\hline
 
Param &  				mean$\pm\sigma$ & 			mean$\pm\sigma$ & 			mean$\pm\sigma$ &				mean$\pm\sigma$		\\
\hline 
$100~\omega_{b }$ & 	$2.243_{-0.02}^{+0.02}$ &		$2.244_{-0.023}^{+0.021}$ &		$2.228^{+0.041}_{-0.041}$&		$2.238_{-0.021}^{+0.02}$\\
$\omega_{cdm }$   & 	$0.1157_{-0.001}^{+0.0012}$ &		$0.1157_{-0.0015}^{+0.0017}$ &	$0.1180^{+0.0029}_{-0.0029}$&	$0.117_{-0.0012}^{+0.0014}$\\
$n_{s }$   & 			$0.9722_{-0.0045}^{+0.0044}$ &	$0.9755_{-0.0059}^{+0.0054}$ &	$0.9690^{+0.0097}_{-0.010}$&		$0.9717_{-0.0051}^{+0.0047}$\\
$\tau_{reio }$  & 		$0.05877_{-0.017}^{+0.0068}$ &	$0.109_{-0.022}^{+0.021}$ &		$0.091^{+0.037}_{-0.037}$&		$0.0901_{-0.017}^{+0.017}$ \\
$H_0$   & 			$69.08_{-0.57}^{+0.49}$ &		$67.73_{-0.77}^{+0.93}$ &		$69.2^{+1.7}_{-1.6}$&			$68.39_{-0.63}^{+0.66}$	\\
$10^{9}A_{s }$  & 		$2.094_{-0.06}^{+0.032}$ &		$2.315_{-0.1}^{+0.085}$ &		$2.24^{+0.16}_{-0.15}$&			$2.237_{-0.072}^{+0.068}$	\\ 
$\sigma_8$  & 			$0.797_{-0.0085}^{+0.0061}$ &		$0.767_{-0.012}^{+0.011}$ &		$0.771^{+0.022}_{-0.022}$&		$0.762_{-0.012}^{+0.01}$\\
$\Omega_{m }$   & 		$0.2909_{-0.006}^{+0.0068}$ &		$0.301^{+0.018}_{-0.016}$ &		$0.293^{+0.016}_{-0.016}$&		$0.2995_{-0.0079}^{+0.0076}$ 	\\
$w_{fld}$&			$-$ &						$-0.958_{-0.042}^{+0.011}$ &		$-1.038^{+0.041}_{-0.060}$&		$-$			\\
$w_{scf}$&			$-$ &						$-$ &						$-$&							$-0.9913_{-0.0087}^{+0.0013}$ \\
$\log{\bar{\xi}}$&		$-$ &						$0.86_{-0.11}^{+0.24}$ &		$1.06^{+0.25}_{-0.28}$&			$-$					\\
$\log{\beta_1}$&		$-$ &						$-$ &						$-$&							$8.6_{-1.7}^{+2.8}$ 	\\	 
$\lambda$&			$-$ &						$-$ &						$-$&							$1.73_{-0.13}^{+0.34}$  	\\ 
\hline

$\chi^2$& 				$1500$ &						$1487$ &						$1488$&						$1478$				\\
$\Delta \chi^2$&		$-$ &						$-13$ &						$-12$&						$-22$				\\
\hline 
 \end{tabular} \\ 
\caption{Constraints of cosmological parameters for $\Lambda$CDM, the T3 model and the ES model obtain using an MCMC and CMB, BAO, JLA and SZ data.}
\label{table:bf-full}
\end{table}

\section{Discussion}
\label{sec:disc}

Interacting dark energy models are discussed in the context of resolving tensions between different data sets and momentum-only models are of particular interest in the context of the $\sigma_8$ tension.
One way of quantifying the tension in $\Lambda$CDM is examining how the best-fit model with the CMB data changes when the fuller data set is included, particularly the Planck SZ (or weak lensing) data. From \autoref{table:bf-cmb}, for $\Lambda$CDM, we see that $\sigma_8 = 0.83 \pm 0.01 $ with the CMB data, which falls to $\sigma_8 = 0.797 \pm 0.007$ when the full data are included.  While the SZ data would prefer even lower values (as will be seen in the momentum-exchange models below), even this compromise model raises the contribution to $\chi^2$ from the CMB significantly ($\Delta \chi^2 \sim 14$, as can be seen in \autoref{table:ch2}).

\begin{figure}[h]
  \centering
  \includegraphics[width=1.0\textwidth]{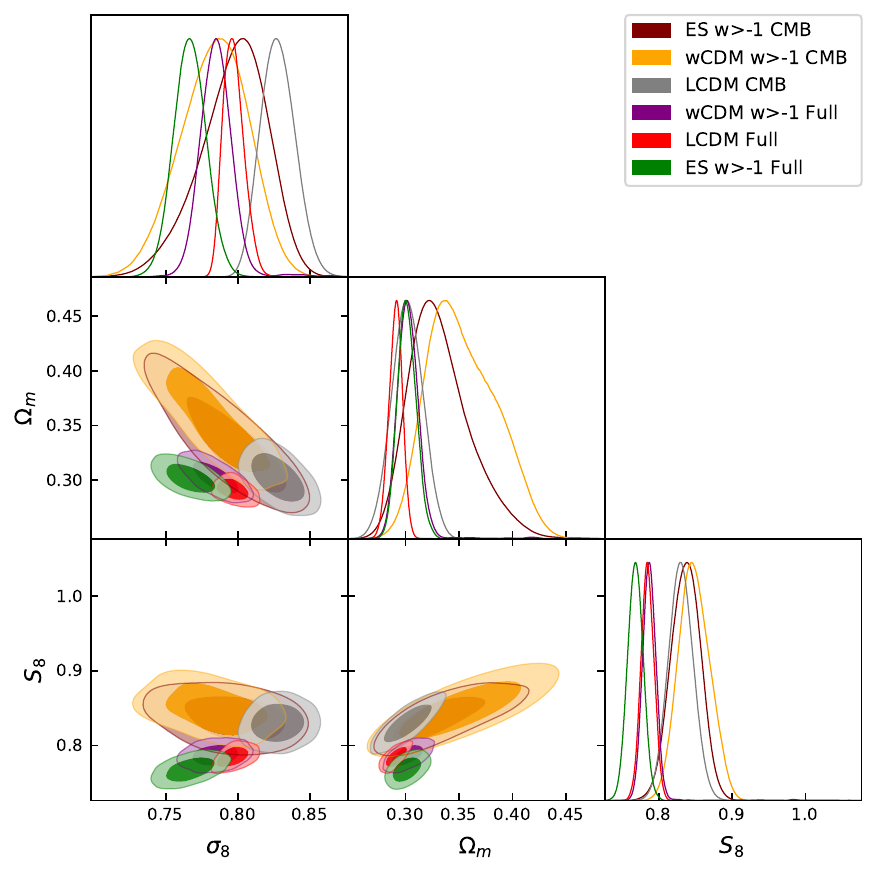}
  \caption{1 and 2 $\sigma$ contours from $\Lambda$CDM  CMB (grey), $\Lambda$CDM CMB, SZ, BAO and JLA (red), ES model with CMB (maroon) and ES model with CMB, SZ, BAO and JLA (red), $w$CDM CMB (orange) and $w$CDM CMB SZ, BAO and JLA  (purple). For this plot we restrict both the ES and $w$CDM models to $w>-1$.} 
\label{fig:ES_S8}
\end{figure}

The situation is improved in the elastic scattering model, but some tension does remain. 
If one looks at the 1D posteriors for $\Omega_M$ and $\sigma_8$ one might naively conclude that the ES model is able to resolve the tension between the CMB data and the SZ data, but if we look at the 2D posterior of $\Omega_M$ and $\sigma_8$, we see there is still very little overlap between the ES CMB contour and the ES CMB+BAO+JLA+SZ contour.  
As above, this can be seen by comparing the CMB contribution to $\chi^2$ for the best-fit models with and without the other data.  In this case, $\Delta \chi^2 \sim 6$, which is a significant improvement over the $\Lambda$CDM case, but is still somewhat concerning.  
This is demonstrated in \autoref{fig:ES_S8};
when we consider the ES model using only the CMB data, the region of the parameter space where the lower values of $\sigma_8$ are found require a very large value for $\Omega_M$ that is not compatible with what we see when we include the other data sets.  This can be seen more clearly when we look at $S_8$, which is motivated by weak lensing and defined as $S_8=\sigma_8\sqrt{\Omega_m/0.3}$.

\begin{figure}[h]
  \centering
  \includegraphics[width=0.9\textwidth]{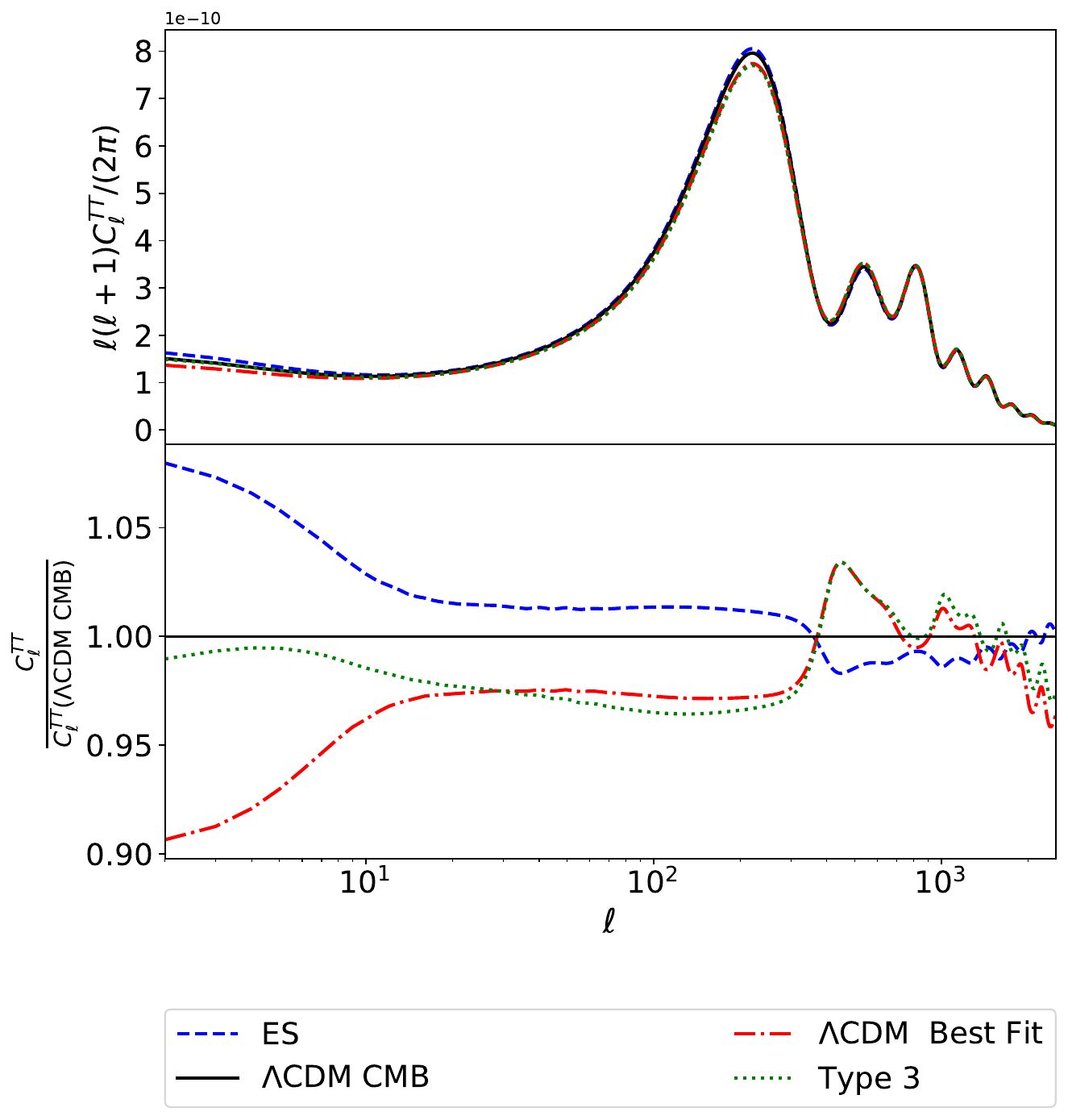}
  \caption{The plot shows the CMB temperature (TT) power spectra when we take the best fit with all of our chosen data for: $\Lambda$CDM, ES model and the Type 3 model, with the $\Lambda$CDM CMB best fit for reference (top) and the ratio of these best fits to $\Lambda$CDM CMB (bottom).
  The ES model has an enhanced ISW contribution which is disfavoured by the data; in the Type 3 case, this effect appears to be compensated somewhat by shifts in the scalar amplitude and $\tau_{reio}$}
\label{fig:BF_CMB}
\end{figure}

The origin of the residual can be found in the fact that the CMB itself prevents the coupling from becoming too large through the late-time ISW effect (see \autoref{fig:CMB_xi} and also \autoref{fig:BF_CMB}.)  This truncates the distribution of the coupling above $\xi \sim 1$, as can be seen in \autoref{fig:ES_MCMC} ; only when the SZ is included does the coupling rise sufficiently (to $\xi \sim 7$) to suppress the power spectrum. 
Despite this, the $\chi^2$ improvement for the elastic scattering models is significant when all the data are included; from \autoref{table:ch2}, we see that it provides a $\Delta \chi^2 \sim 15$, which is sufficient to justify the extra two degrees of freedom.  

\begin{figure}[h]
  \centering
  \includegraphics[width=1.0\textwidth]{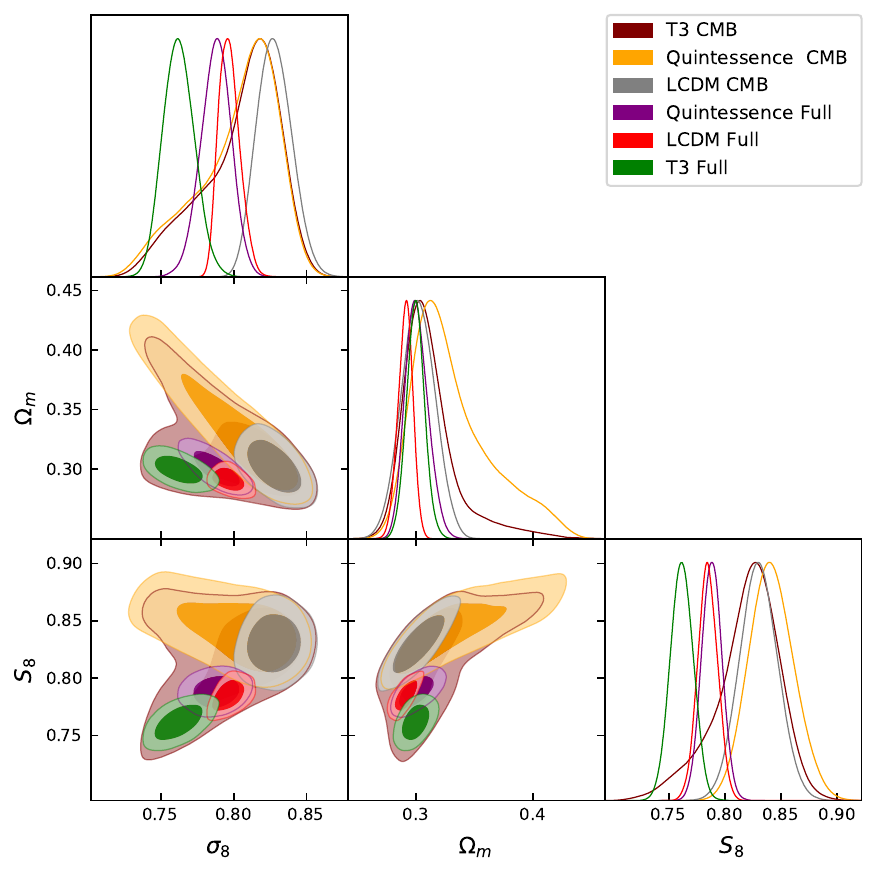}
  \caption{1 and 2 $\sigma$ contours from $\Lambda$CDM  CMB (grey), $\Lambda$CDM CMB, SZ, BAO and JLA (red), T3 model with CMB (maroon) and T3 model with CMB, SZ, BAO and JLA (red), Quintessence CMB (orange) and Quintessence CMB SZ, BAO and JLA (purple).
  } 
\label{fig:T3_S8}
\end{figure}

The \emph{Type 3} contours in the $S_8-\Omega_m$ plane are very similar to what we see in the ES case when we consider all data (see \autoref{fig:T3_S8}).
There is however a significant difference in the CMB-only results; the \emph{Type 3} models are looser and seem to avoid the tension entirely.  In this case, it appears that the CMB constraint on the coupling arising from the ISW contribution is not as strong, meaning some significant suppression of the power spectrum can be achieved without disrupting the fit to the CMB data (see \autoref{fig:BF_CMB}.) 
Therefore the \emph{Type 3} models have a significantly lower chi-squared still, $\Delta \chi^2 \sim 23$.  Again, this is strong enough to justify the extra degrees of freedom, making these models a better candidate to resolve the tensions.   

When not constrained by the model, the SZ data prefers a lower value for $\sigma_8 = 0.76 \pm 0.01$
We note that the chi-squared differences largely come from the CMB and SZ data.  As the background evolution in all models are consistent with a cosmological constant, the JLA and BAO data do not distinguish between the models. 

There is also a tension between the Planck CMB observations and the weak lensing observations of KiDS \cite{Hildebrandt:2016iqg,Asgari:2020wuj} and DES \cite{Abbott:2017wau}). However, the analysis of weak lensing data requires very careful consideration as it heavily depends on nonlinear effects, which are model dependent (see e.g. \cite{SpurioMancini:2019rxy,Schneider:2019xpf,Martinelli:2020yto,Troster:2020kai,Bose:2021mkz}). A complete reanalysis of weak lensing data for interacting dark energy models is beyond the scope of this paper but is the subject of future work.

\begin{table}

    \centering
    \begin{tabular}{|l|c|c|c|c|c|} 
    \hline
	&		&	$\Lambda$CDM	&	ES $w>-1$	&	ES $w<-1$	&	T3	\\
	\hline
CMB	&	$\chi^2$	& $	805.1	$ & $	797.4	$ & $	800.8	$ & $	791.4	$ \\
	&	$\Delta$ $\Lambda$CDM	& $	-	$ & $	7.7	$ & $	4.3	$ & $	13.7	$ \\
\hline
BAO	&	$\chi^2$	& $	2.5	$ & $	3.2	$ & $	2.6	$ & $	2.7	$ \\
	&	$\Delta$ $\Lambda$CDM	& $	-	$ & $	-0.7	$ & $	-0.1	$ & $	-0.2	$ \\
\hline
JLA	&	$\chi^2$	& $	683.3	$ & $	683.6	$ & $	683.4	$ & $	683.3	$ \\
	&	$\Delta$ $\Lambda$CDM	& $	-	$ & $	-0.3	$ & $	-0.1	$ & $	-0.06	$ \\
\hline
SZ	&	$\chi^2$	& $	9.5	$ & $	1.5	$ & $	1.0	$ & $	0.3	$ \\
	&	$\Delta$ $\Lambda$CDM	& $	-	$ & $	8.0	$ & $	8.5	$ & $	9.2	$ \\
\hline
Total	&	$\chi^2$	& $	1500.4	$ & $	1485.7	$ & $	1487.8	$ & $	1477.7	$ \\
	&	$\Delta$ $\Lambda$CDM	& $	-	$ & $	14.7	$ & $	12.6	$ & $	22.7	$ \\
\hline
\end{tabular}
\caption{Break down of the contributions to the $\chi^2$, for the best fit value, when all the data is considered, for $\Lambda$CDM, ES $w>-1$, ES $w<-1$ and T3 model.}
\label{table:ch2}
\end{table}

\section{Conclusions} 
\label{sec:concl}
In this paper we have explored models of interacting dark energy with pure momentum exchange. We have tested their viability when they are compared to data from early and late time cosmological probes using a modified version of \CLASS and \MontePython. We found that they are competitive with the cosmological standard model $\Lambda$CDM, even obtaining significantly better fits when certain data sets are included.
However, we should note that the best way to compare different models is using the Bayesian evidence approach. We will employ a full model selection
analysis based on the Bayesian evidence in future work.

Looking to the future, there is still much to learn about interacting dark energy models. Current and forthcoming experiments such as the Dark Energy Spectroscopic Instrument (DESI)\footnote{\url{https://www.desi.lbl.gov/}}~\cite{Aghamousa:2016zmz}, 
Euclid\footnote{\url{http://euclid-ec.org}}~\cite{Blanchard:2019oqi,Laureijs:2011gra}, the Nancy Grace Roman Space Telescope\footnote{\url{https://www.nasa.gov/roman}}~\cite{spergel2015wide}, and the Vera C. Rubin Observatory’s Legacy Survey of Space and Time (LSST)\footnote{\url{https://www.lsst.org/}}~\cite{Mandelbaum:2018ouv}  will test these models in greater detail and forecasting this is a valuable process (see e.g. \cite{Amendola_2012, Figueruelo:2021elm,Carrilho_2021}). We only consider models with time-independent couplings in this paper, but there is no obvious physical reason why this would be the case. In addition to studying models with a parameterised coupling (possibly similar to a $w_0$, $w_a$ parameterisation commonly used to study a dynamical equation of state), one could employ reconstruction techniques such as the ones used in \cite{Hogg:2020rdp, Crittenden:2011aa}, to explore how the coupling could evolve with time in a very general way. Finally, there has been little work looking at the micro-physics of IDE models. In general one would expect the introduction of the coupling to DM to introduce large corrections to the DE field, possibly disrupting its ability to cause the accelerated expansion. Momentum only models would likely not have this problem and investigating this in a general and rigorous way is the subject of future work.

\section{Acknowledgements} 

ML's research is supported by an STFC studentship. AP is a UK Research and Innovation Future Leaders Fellow, grant MR/S016066/1. RC is supported by STFC grant ST/S000550/1. For the purpose of open access, the author has applied a Creative Commons Attribution (CC BY) licence to any Author Accepted Manuscript version arising.  Supporting research data are available on reasonable request from the authors.  We would like to thank  Benjamin Bose for the use of his data; Michaela Lawrence and Alessio Spurio Mancini for useful discussions; and Jascha Schewtschenko and Gary Burton for their invaluable technical support.
This project has made use of the SCIAMA High Performance Computing cluster at the ICG.

{Whilst this paper was being finalised, the following paper that covers similar topics was added to the arXiv \cite{Jimenez:2021ybe}.}

\bibliographystyle{apsrev}
\bibliography{references.bib}
\end{document}